\documentclass[fleqn,12pt,epsfig
    ,final            % use final for the camera ready runs 
    ,numberedheadings % uncomment this option for numbered sections
    ,cmfonts          % add further options here if necessary
  ]
  {aipproc}
\usepackage{young}
\usepackage{graphicx}
\layoutstyle{6x9}
\def\1{\mbox{l\hspace{-0.53em}1}}
\begin{document}

\title{The Charge Conjugation Quantum Number in Multiquark Systems}


\author{Fl. Stancu}{
  address={Universit\'e de Li\`ege, Institut de Physique B5, Sart Tilman, B-4000 Li\`ege 1, Belgium\\
E-mail: fstancu@ulg.ac.be}
\footnote{Based on a talk given at the Joint Meeting Heidelberg-Li\`ege-
Paris-Wroclaw (HLPW08): Three Days of Strong Interactions and Astrophysics, 
Spa, Li\`ege, March 6-8, 2008}}

\begin{abstract}

We discuss the charge conjugation quantum number for tetraquarks and
meson-meson molecules, both seen as possible interpretations of the
newly found $XYZ$ charmonium-like resonances.

\end{abstract}

\maketitle

\section{Introduction}

The discovery by Belle \cite{Choi:2003ue}
of  the very narrow narrow $X(3872)$ resonance
has revived the 
interest in heavy quarkonium, both experimentally and theoretically.  
The observation of $X(3872)$ has been confirmed by 
CDF \cite{Acosta:2003zx}, D0 \cite{Abazov:2004kp} and Babar 
Collaborations \cite{Aubert:2004ns}. 
Other charmonium-like states,
as for example 
$X(3940)$ \cite{Abe:2004zs}, $Y(4260)$ \cite{Aubert:2005rm}
and $Z^+(4430)$ \cite{:2007wga} 
have been uncovered in B-factory experiments. 
A partial list is shown in Table \ref{exotics}. 
%The only charged resonance is
%$Z^+(4430)$. 
Several review papers,
as for example \cite{Swanson,Godfrey:2008nc},
discuss the difficulty of interpreting
these resonances as charmonium states. Seen as 
exotics, they can possibly be
tetraquarks, meson-meson molecules, hybrids, glueballs, etc. 

\begin{table}[h!]
\caption{Properties of newly discovered charmonium-like resonances}
\label{exotics}
\renewcommand{\arraystretch}{1.25}
\begin{tabular}{|c|c|c|c|c|}
\hline%\hline
Resonance & Mass & Width & J$^{PC}$ & Decay modes \\
          & (MeV)& (MeV) & \\
\hline
X(3872)   &  3871.4 $\pm$ 0.6   &  $< 2.3$  & 1$^{++}$ & 
$ \pi^+ \pi^- J/\Psi, \gamma J/\Psi $   \\
%X(3875)   &  3875.5 $\pm$ 1.5 & 3.0$^{+2.1}_{-1.7}$ & & 
%$  D^0{\bar D}^0 \pi^0$ \\
X(3940)  &  3942 $\pm$ 9     &  37 $\pm$ 17 & J$^{P+}$ &
$  D{\bar D}^*$ \\
Y(3940)  &  3943 $\pm$ 17    &  87 $\pm$ 
 34  & J$^{P+}$ &
$\omega J/\Psi$ \\ 
Z(3930)  &  3929 $\pm$ 5     &  29 $\pm$ 10  & 2$^{++}$ &
$  D{\bar D}$ \\
X(4160)  & 4156 $\pm$ 29     &  139$^{+113}_{-65}$   &  J$^{P+}$ &
$  D^*{\bar D}^*$ \\
Y(4260)   &  4264$\pm$ 12    &  83 $\pm$ 22 & 1$^{--}$ & 
$ \pi^+ \pi^- J/\Psi $ \\
Z$^{+}$(4430)  &  4433 $\pm$ 5     &   45 $^{+35}_{-18}$ & ? &
$ \pi^{\pm} \Psi'$   \\
Y(4660)  & 4664 $\pm$ 12     &   48 $\pm$ 15 & 1$^{--}$ & 
$ \pi^+ \pi^- \Psi'$ \\
\hline  %\hline
\end{tabular}
\end{table}

Whatever picture is adopted, it is important to properly determine
the charge conjugation number $C$ of the resonance. From Table \ref{exotics} one 
can see that $C$ is known in nearly all cases. Here we shall
discuss the tetraquark option, for example  \cite{Hogaasen:2005jv}, 
and the molecular option, for example \cite{Voloshin:2003nt} or 
\cite{Liu:2008fh}, where we noticed
that some difficulty has been encountered in defining it.

%%%%%%%%%%%%%%%%%%%%%%%%%%%%%%%%%%%%%%%%%%%%%%%%%%%%%%%%%%%%%%%%%%%
\section{The basis states}

We recall the definition  of tetraquark basis states as given 
in Refs.   \cite{Brink:1994ic,Brink:1998as,Stancu:1991rc}. This basis 
can provide a direct connection with the diquark-antidiquark states 
\cite{Maiani:2004vq} and also with the molecular states 
\cite{Voloshin:2003nt,Liu:2008fh,TORNQVIST}.

\begin{figure}[h!]
\label{fig1}
%\begin{center}
\includegraphics*[width=10.0cm,keepaspectratio]{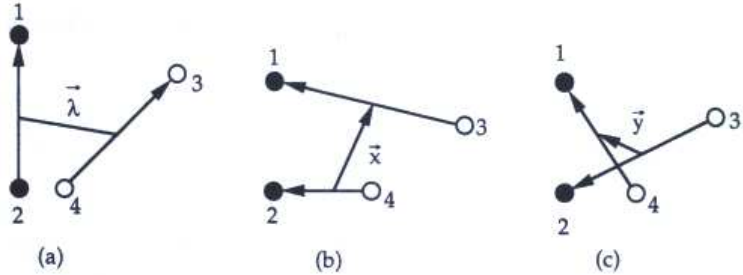}
%\end{center} 
\caption{Three independent relative coordinate systems. Solid and 
open circles represent quarks and antiquarks respectively: (a)
diquark-antidiquark channel, (b) direct meson-meson channel, (c) exchange
meson-meson channel. }
\end{figure}

In Fig. \ref{fig1} we suppose that the particle 1 is the charmed quark $c$, 
the particle 2  a light quark $q = u,d$, particle 3 the anticharmed quark
$\bar c$ and particle 4 a light antiquark $\bar q = {\bar u}, {\bar d}$.

The total wave function of a tetraquark is a linear combination of products of
orbital, spin, flavour and colour parts.
We shall successively introduce the orbital, colour and spin parts
of the wave function. The flavor part is fixed.
%%%%%%%%%%%%%%%%%%%%%%%%%%%%%%%%%%%%%%%%%%%%%%%%%%%%%%%%%

\subsection{The orbital part} 
There are at least three possible ways to define the relative
coordinates. The three relevant possibilities for our problem are
shown in Fig. 1. 
In the cases $(a),(b)$ and $(c)$ the internal coordinates are 
\begin{equation}
   \vec{\sigma}  = \frac{1}{\sqrt{2}} (\vec{r_1}-\vec{r_2}),
~~~\vec{\sigma'} = \frac{1}{\sqrt{2}} (\vec{r_3}-\vec{r_4}),
~~~\vec{\lambda} = \frac{1}{2}(\vec{r_1}+\vec{r_2}-\vec{r_3}-\vec{r_4}),   
\end{equation}
\begin{equation}\label{direct}
   \vec{\rho}  = \frac{1}{\sqrt{2}} (\vec{r_1}-\vec{r_3}), 
~~~\vec{\rho'} = \frac{1}{\sqrt{2}} (\vec{r_2}-\vec{r_4}),
~~~\vec{x}     = \frac{1}{2}(\vec{r_1}-\vec{r_2}+\vec{r_3}-\vec{r_4}),   
\end{equation}
\begin{equation}\label{exchange}
   \vec{\alpha}  = \frac{1}{\sqrt{2}} (\vec{r_1}-\vec{r_4}), 
~~~\vec{\alpha'} = \frac{1}{\sqrt{2}} (\vec{r_2}-\vec{r_3}),
~~~\vec{x}     = \frac{1}{2}(\vec{r_1}-\vec{r_2}-\vec{r_3}+\vec{r_4}).   
\end{equation}
The first system of coordinates is convenient when the quarks or 
antiquarks are correlated to form diquarks, as  
in the diquark-antidiquark model. 
%of Ref. \cite{Maiani:2004vq}. 
The coordinates (2) and (3) 
called direct and exchange meson-meson 
channels, are useful in describing  strong decays or introduce
a molecular picture  \cite{TORNQVIST}, provided a
quark structure is imposed at short separations.  One should
use the system which is more appropriate for a given problem. But in 
specific calculations one can pass from one coordinate system to the other
by orthogonal transformations \cite{Brink:1998as}.

%%%%%%%%%%%%%%%%%%%%%%%%%%%%%%%%%%%%%%%%%%%%%%%%%%%%%%%%%%%%%%%%%%

\subsection{The colour part}
In the colour space one can construct a colour singlet tetraquark 
state using intermediate couplings associated to the three
coordinate systems defined above. In this way one obtains three 
distinct bases  \cite{Stancu:1991rc}
\begin{equation}\label{diquark}
|\overline{3}_{12} 3_{34} \rangle, ~~~ |6_{12} \overline{6}_{34} \rangle,
\end{equation}
\begin{equation}\label{directcolour}
|1_{13} 1_{24} \rangle, ~~~ |8_{13} 8_{24} \rangle,
\end{equation} 
\begin{equation}\label{exchangecolour}
|1_{14} 1_{23} \rangle, ~~~ |8_{14} 8_{23} \rangle.
\end{equation} 
In the basis (\ref{diquark})
the states 3 and $\overline{3}$ are antisymmetric and 6 and $\overline{6}$
are symmetric under interchange of quarks or antiquarks. This basis
is convenient in a diquark-antidiquark picture.
In Ref. \cite{Maiani:2004vq} only the  $|\overline{3}_{12} 3_{34} \rangle$
state has been considered which restricts the spectrum to
half of the allowed states  \cite{Stancu:2006st}.
The sets   (\ref{directcolour}) or  (\ref{exchangecolour}), 
contain a singlet-singlet colour
and an octet-octet colour state. The amplitude of the latter
vanishes asymptotically when a tetraquark 
decays into two mesons. These
are called \emph{hidden colour} states, by analogy to states
which appear in  the nucleon-nucleon problem. 
When the amplitude of a \emph{hidden colour} state is far the most dominant
in the wave function, the corresponding open channel acquires, in exchange, 
a tiny value for its amplitude  and in this way the authors of 
Ref.  \cite{Hogaasen:2005jv} explained the small width of $X(3872)$,
when interpreted as a $c \bar c q \bar q$ tetraquark.
%%%%%%%%%%%%%%%%%%%%%%%%%%%%%%%%%%%%%%%%%%%%%%%%%%%%%%%%%%%%%%%%

\subsection{The spin part} 

As the quarks and antiquarks are spin 1/2 particles, the total 
spin of a tetraquark can be $S = 0, 1$ or 2. This can be obtained
by first coupling two particles to each other and then couple
the two subsystems together. Let us denote the intermediate 
coupling states 
between two quarks (antiquarks) by $S_{ij}$ for spin 0 (Scalar) and 
by $A_{ij}$ for spin 1 (Axial). For a quark-antiquark pair we denote  the 
states by  $P_{ij}$ for spin 0 (Pseudoscalar) and by $V_{ij}$ for spin 1
(Vector).
We need their permutation symmetry properties under a given 
transposition $(ij)$.
For quark (antiquark) pairs we have
\begin{equation}\label{transp12}
    (12) |S_{12} \rangle = - |S_{12} \rangle,
~~~ (12) |A_{12} \rangle = + |A_{12} \rangle,
\end{equation}
and 
\begin{equation}\label{transp34}
    (34) |S_{34} \rangle = - |S_{34} \rangle,
~~~ (34) |A_{34} \rangle = + |A_{34} \rangle,
\end{equation}
For quark-antiquark pairs we have
\begin{equation}\label{transp13}
    (13) |P_{13} \rangle = - |P_{13} \rangle,
~~~ (13) |V_{13} \rangle = + |V_{13} \rangle,
\end{equation}
and  
\begin{equation}\label{transp24}
    (24) |P_{24} \rangle = - |P_{24} \rangle,
~~~ (24) |V_{24} \rangle = + |V_{24} \rangle.
\end{equation}
 
For $S = 0$ there are two independent basis states for each channel.
The bases associated to 
(\ref{diquark}), (\ref{directcolour}) and  (\ref{exchangecolour}) are
\begin{equation}\label{diquark0}
|A_{12} A_{34} \rangle, ~~~ |S_{12} S_{34} \rangle,  
\end{equation}
\begin{equation}\label{directcolour0}
|P_{13} P_{24} \rangle,~~~ |(V_{13} V_{24})_0 \rangle, 
\end{equation} 
\begin{equation}\label{exchangecolour0}
|P_{14} P_{23} \rangle,~~~ |(V_{14} V_{23})_0 \rangle, 
\end{equation} 

For $S = 1$ there are three independent states
in each channel, to be identified by three
distinct Young tableaux. 
As an example we give the basis 
for the direct meson-meson channel  \cite{Brink:1998as}
\begin{equation}\label{directcolour1}
|(P_{13} V_{24})_1 \rangle, ~~~~|(V_{13} P_{24})_1 \rangle, 
~~~|(V_{13} V_{24})_1 \rangle.
\end{equation}
The lower index indicates the total spin 1.

The case $S = 2$ is trivial. There is a single state 

\begin{equation}\label{SPIN2}
\chi^S = |(V_{13} V_{24})_2 \rangle , 
\end{equation}
which is symmetric under any permutation of particles.
%%%%%%%%%%%%%%%%%%%%%%%%%%%%%%%%%%%%%%%%%%%%%%%%%%%%%%%%%%%%

\section{Charge conjugation}
We deal with the ground state, {\emph i.e.} we have    $J = S$.
Making the identification 1 = $c$, 2 = $q$, 3 = $\overline c$ and 
4 = $\overline q$, introduced above, it is convenient to first couple 
1 to 3 and 2 to 4 and then the subsystem 13 to 24, as in Fig. 1b. 
The charge conjugation is equivalent
to applying the permutation (13)(24) to the wave function. 

\subsection{Tetraquarks}

The colour state (\ref{directcolour}) does not change under the 
permutation (13)(24). The discussion holds for spin states only.

From the properties (\ref{transp13}) and (\ref{transp24}) 
one can see that the spin states (\ref{directcolour0}) and
(\ref{SPIN2}) remain unchanged under the permutation (13)(24). 
It follows that the 
states $J^P=0^+$ and $J^P=2^+$ have charge conjugation $C = +$.

For $J^P=1^+$ the situation is slightly more complicated. 
There are six linearly independent basis 
vectors built as products of colour (\ref{directcolour}) and 
spin (\ref{directcolour1}) states.
\begin{eqnarray}\label{eq:alphai}
\hskip -10pt& \alpha_1 =
%(q_1 \overline{q}_3)^1_0 \otimes (q_2 \overline{q}_4 )^{{1}}_{1},\
| 1_{13} 1_{24} (P_{13} V_{24})_1 \rangle,\ 
&  \alpha_2 =
%(q_1 \overline{q}_3)^1_1 \otimes (q_2 \overline{q}_4)^{{1}}_{0},\nonumber\\
%
| 1_{13} 1_{24} (V_{13} P_{24})_1 \rangle,\nonumber\\
\hskip -10pt& \alpha_3 =
%(q_1 \overline{q}_3)^1_1 \otimes (q_2\overline{q}_4 )^{{1}}_{1},\ 
| 1_{13} 1_{24}~ (V_{13} V_{24})_1 \rangle, \  
&\alpha_4 =
%(q_1 \overline{q}_3)^8_0 \otimes (q_2 \overline{q}_4)^{{8}}_{1}, \\ 
%
| 8_{13} 8_{24} (P_{13} V_{24})_1 \rangle, \nonumber\\
\hskip -10pt& \alpha_5 =
%(q_1 \overline{q}_3)^8_1 \otimes (q_2\overline{q}_4 )^{{8}}_{0}, \
| 8_{13} 8_{24} (V_{13} P_{24})_1 \rangle, \ 
& \alpha_6 =
%(q_1 \overline{q}_3)^8_1 \otimes  (q_2 \overline{q}_4)^{{8}}_{1} . \nonumber
| 8_{13} 8_{24}~ (V_{13} V_{24})_1 \rangle.
\end{eqnarray}
Under the permutation (13)(24) 
the basis vectors   $|\alpha_1 \rangle$, $|\alpha_2 \rangle$,
$|\alpha_4 \rangle$, $|\alpha_5 \rangle$ change sign thus have charge 
conjugation $C = - $. On the other hand
$|\alpha_3 \rangle$ and $| \alpha_6 \rangle $ do not change sign, thus
have charge conjugation $C = + $. Also, by construction, they correspond
to tetraquarks where the spin of the $c \bar c$ pair is $S_{c \bar c} = 1$. 

All states $\alpha_i$ of Eq. (\ref{eq:alphai}) used to construct Table 1 
of Ref. \cite{Hogaasen:2005jv} have inadvertently
been associated to $ C = + $, as explained in  Ref. \cite{Stancu:2006st}.

%%%%%%%%%%%%%%%%%%%%%%%%%%%%%%%%%%%%%%%%%%%%%%%%%%%%%%%%%%%%%%

\subsection{Meson-meson molecules}
The only interesting case is $J^P = 1^+$. Here
we  need the basis states in the exchange channel 
corresponding to Fig. 1c.
As for the direct channel there are six linearly independent 
basis vectors

\begin{eqnarray}\label{eq:betai}
\hskip -10pt& \beta_1 =
| 1_{14} 1_{23} (P_{14} V_{23})_1 \rangle,\ 
&  \beta_2 =
| 1_{14} 1_{23} (V_{14} P_{23})_1 \rangle,\nonumber\\
\hskip -10pt& \beta_3 =
| 1_{14} 1_{23}~ (V_{14} V_{23})_1 \rangle, \  
&\beta_4 =
| 8_{14} 8_{23} (P_{14} V_{23})_1 \rangle, \nonumber\\
\hskip -10pt& \beta_5 =
| 8_{14} 8_{23} (V_{14} P_{23})_1 \rangle, \ 
& \beta_6 =
| 8_{14} 8_{23}~ (V_{14} V_{23})_1 \rangle.
\end{eqnarray}
Using the Appendix C of Ref. \cite{Brink:1998as} 
and the relations  \cite{Stancu:1991rc}
\begin{equation} 
|1_{14} 1_{23} \rangle = \frac{1}{3}|1_{13} 1_{24} \rangle 
+  \frac{2 \sqrt{2}}{3}|8_{13} 1_{24} \rangle,  \nonumber \\
~~~|8_{14} 8_{23} \rangle = \frac{2 \sqrt{2}}{3}|1_{13} 1_{24} \rangle 
-  \frac{1}{3} |8_{13} 1_{24} \rangle,
\end{equation}
one can express $\beta_i$ in terms of $\alpha_i$ or vice-versa.
In particular, one has 
\begin{equation}
\beta_1= \frac{1}{6}(\alpha_1 + \alpha_2) - \frac{1}{3\sqrt{2}} \alpha_3
+ \frac{\sqrt{2}}{3}(\alpha_4 + \alpha_5) - \frac{2}{3} \alpha_6,
\end{equation} 
\begin{equation}
\beta_2= \frac{1}{6}(\alpha_1 + \alpha_2) + \frac{1}{3\sqrt{2}} \alpha_3
+ \frac{\sqrt{2}}{3}(\alpha_4 + \alpha_5) + \frac{2}{3} \alpha_6,
\end{equation}
\begin{equation}
\beta_3= - \frac{1}{3\sqrt{2}} (\alpha_1 - \alpha_2) 
- \frac{2}{3}(\alpha_4 - \alpha_5)
\end{equation}
\begin{equation}
\beta_4= \frac{\sqrt{2}}{3}(\alpha_1 + \alpha_2) - \frac{1}{6}(\alpha_4 + \alpha_5)
- \frac{2}{3}\alpha_3 + \frac{1}{3\sqrt{2}} \alpha_6,
\end{equation}
\begin{equation}
\beta_5= \frac{\sqrt{2}}{3}(\alpha_1 + \alpha_2) - \frac{1}{6}(\alpha_4 + \alpha_5)
+ \frac{2}{3}\alpha_3 - \frac{1}{3\sqrt{2}} \alpha_6,
\end{equation}
\begin{equation}
\beta_6= - \frac{2}{3}(\alpha_1 - \alpha_2) + \frac{1}{3\sqrt{2}} (\alpha_4 - \alpha_5)
\end{equation}
From the properties of $\alpha_i$ one can infer  that $\beta_1$,  
$\beta_2$,  $\beta_4$ and $\beta_5$
do not have a definite charge conjugation. Contrary, $\beta_3$ and  $\beta_6$
have $C = -$, because they are linear combinations of states with  $C = -$.
Inverting the above relations one can obtain $\alpha_i$ in terms of
$\beta_i$. We have, for example
\begin{equation}
\alpha_3 =  - \frac{\sqrt{2}}{6} (\beta_1 - \beta_2)
- \frac{2}{3} (\beta_4 - \beta_5), 
~~~~ \alpha_6 = \frac{2}{3}(\beta_1 - \beta_2)
 + \frac{\sqrt{2}}{6} (\beta_4 - \beta_5).
\end{equation}
Section 3.1 implies that both $\beta_1 - \beta_2$ and $\beta_4 - \beta_5$
have $C = +$. From the identification 
1 = $c$, 2 = $q$, 3 = $\overline c$ and 4 = $\overline q$ 
and the definitions (\ref{eq:betai})
it follows that 
\begin{equation}
\beta_1 = D^0 \overline{D}^{*0}, ~~~\beta_2 = D^{*0} \overline{D}^{0}.
\end{equation}
Thus one has
\begin{equation}
\alpha_3 = - \frac{\sqrt{2}}{6} 
(D^0 \overline{D}^{*0} - D^{*0} \overline{D}^{0})
- \frac{2}{3} (\beta_4 - \beta_5)
\end{equation}
\begin{equation}
\alpha_6 = -\frac{2}{3} (D^0 \overline{D}^{*0} - D^{*0} \overline{D}^{0})
+ \frac{\sqrt{2}}{6}(\beta_4 - \beta_5),
\end{equation}
where the second component is a \emph{hidden colour} state with 
the same spin structure as $(D^0 \overline{D}^{*0} - D^{*0} \overline{D}^{0})$.
It follows that a
molecular structure with $C = +$  can be obtained only from the 
difference  $\beta_1 - \beta_2$ plus its \emph{hidden colour}  counterpart.
Except for this part, 
our result is in agreement with the discussion given in 
Ref. \cite{Liu:2008fh},
but in disagreement with the  molecular interpretation of 
Ref.  \cite{Voloshin:2003nt} .
The \emph{ hidden colour} component with $C = +$ can influence the energy 
of the system at short separations \cite{Brink:1994ic}. Such contributions 
have been ignored in the simple molecular picture, 
because the molecules are point-like particles
\cite{Voloshin:2003nt,Liu:2008fh,TORNQVIST}.

One can also find that
\begin{equation}
\alpha_1 + \alpha_2 = \beta_1 + \beta_2, 
~~~~ \alpha_5 + \alpha_6 = \beta_4 + \beta_5
 \end{equation}
which show that  $\beta_1 + \beta_2$  and
$\beta_4 +  \beta_5$  have  $C = -$. 
Thus the (open channel) molecular state with $C = - $ reads
\begin{equation}\label{OPEN}
\beta_1 + \beta_2 =  D^0 \overline{D}^{*0} + D^{*0} \overline{D}^{0}
 \end{equation}  
again in agreement with Ref. \cite{Liu:2008fh}. Interestingly, its hidden channel 
counterpart, $\beta_4 +  \beta_5$ is completely decoupled from the open channel
(\ref{OPEN}).

\vspace{1cm}
{\bf Acknowledgements} I am grateful to Yoshi Fujiwara for pointiong out
an anomaly in the relations presently numbered as Eq. (18).

%%%%%%%%%%%%%%%%%%%%%%%%%%%%%%%%%%%%%%%%%%%%%%%%%%%%%%%%%%%%

%%%%%%%%%%%%%%%%%%%%%%%%%%%%%%%%%%%%%%%%%%%%%%%%%%%%%%%%%%
\end{document}